\documentclass[article]{emulateapj}
\usepackage{graphicx}
\usepackage{natbib} 
\usepackage{array}
\usepackage{times}
\def\apjl{{ApJ}}                
\def\apjs{ {ApJS}}               
\def\ao{ {Appl.~Opt.}}           
             
\def\aap{ {A\&A}}

\def\arcsec{$^{\prime\prime}$}

\newcommand{\be}{\begin{equation}}
\newcommand{\ee}{\end{equation}}

\newcommand{\msun}{{$M_{\odot}$}}
\newcommand{\mstar}{{$M_{\star}$}}

\newcommand{\se}{s$^{-1}$ }
 
\newcommand{\gtsima}{$\; \buildrel > \over \sim \;$}
\newcommand{\ltsima}{$\; \buildrel < \over \sim \;$}
\newcommand{\prosima}{$\; \buildrel \propto \over \sim \;$}
\newcommand{\gsim}{\lower.5ex\hbox{\gtsima}}
\newcommand{\lsim}{\lower.5ex\hbox{\ltsima}}
\newcommand{\simgt}{\lower.5ex\hbox{\gtsima}}
\newcommand{\simlt}{\lower.5ex\hbox{\ltsima}}
\newcommand{\simpr}{\lower.5ex\hbox{\prosima}}
\newcommand{\es}{erg~s$^{-1}$}

\newcommand{\cxo}{\textit{Chandra}}
\newcommand{\hst}{\textit{Hubble Space Telescope}}

\newcommand{\lx}{$L_{\rm X}$}

\def\ellx{${\ell}_{\rm X}$}

\begin{document}

\title{Sub-Eddington super-massive black hole activity in Fornax early type galaxies}

\author{Nathan Lee\altaffilmark{1}, Elena Gallo\altaffilmark{1}, Edmund Hodges-Kluck\altaffilmark{1,2}, Patrick Cot\'e\altaffilmark{3}, Laura Ferrarese\altaffilmark{3}, Brendan Miller\altaffilmark{4},Vivienne Baldassare\altaffilmark{5,6}, Richard Plotkin\altaffilmark{7}, Tommaso Treu\altaffilmark{8} }
\altaffiltext{1}{Department of Astronomy, University of Michigan, 1085 S. University, Ann Arbor, MI 48109}
\altaffiltext{2}{NASA Goddard Space Flight Center, 8800 Greenbelt Rd, Greenbelt, MD 20771}
\altaffiltext{3}{National Research Council of Canada, Herzberg Astronomy and Astrophysics Research Centre, 5071 West Saanich Road, Victoria, Canada}
\altaffiltext{4}{School of Physical Sciences, College of St. Scholastica, The College of St. Scholastica, 1200 Kenwood Avenue, Duluth, MN 55811}
\altaffiltext{5}{Department of Astronomy, Yale University, New Haven, CT 06520}
\altaffiltext{6}{Einstein Fellow}
\altaffiltext{7}{International Centre for Radio Astronomy Research, Curtin University, GPO Box U1987, Perth, WA 6845, Australia}
\altaffiltext{8}{Department of Physics and Astronomy, University of California, Los Angeles, 475 Portola Plaza, Los Angeles, CA 90095}

\begin{abstract}
We characterize the incidence and intensity of low-level super-massive black hole activity within the Fornax cluster, through X-ray observations of the nuclei of 29 quiescent early-type galaxies. Using the \textit{Chandra X-ray Telescope}, we target 17 galaxies from the \hst\ Fornax Cluster Survey, down to a uniform (3$\sigma$) limiting \textbf{X-ray} luminosity threshold of $5\cdot10^{38}$ergs$^{-1}$, which we combine with deeper, archival observations for an additional 12 galaxies. A nuclear X-ray point-source is detected in 11 out of 29 targets. After accounting for the low mass X-ray binary contamination to the nuclear X-ray signal, the X-ray active fraction is measured at $26.6\% \pm 9.6\%$.
The results from this analysis are compared to similar investigations targeting quiescent early types in the Virgo cluster, as well as the field.  After  correcting  for  the  different  mass  distributions, the measured Fornax active fraction is less than the field fraction, at more than 3$\sigma$, confirming that the funneling of gas to the nuclear regions of cluster members is inhibited compared to those galaxies in the field. At the same time, we find no statistically significant difference between Fornax and Virgo galaxies, with only marginal evidence for a lower active fraction in Fornax (1 $\sigma$); if real, owing to Fornax's higher galaxy number density, this could indicate that galaxy-galaxy interactions are more effective at gas removal than galaxy-gas effects. 
\end{abstract}



\section{Introduction} \label{sec:intro}
\par The growth of super-massive black holes (SMBHs) appears to be anti-hierarchical, in the sense that active accretion is concentrated in higher and lower mass SMBHs at earlier and later cosmological times, respectively (e.g., \citealt{heckman04,shankar09,bongiorno12}). The peak of the quasar space densities around z$\simeq$2 (e.g., \cite{hasinger05}), and the $\simlt10^8$ yr quasar lifetimes \citep{Yu&Tremaine}, imply that ``inactive" galactic nuclei host SMBHs accreting at highly sub-Eddington levels in a post-quasar phase (e.g., \citealt{soltan,croton06,hopkins07}). While several studies (e.g., \citealt{pellegrini05,soria06,boroson11}) have provided a detailed census of low-level SMBH ativity within the local universe, including as a function of the host galaxy properties \citep{pellegrini10}, {the impact of environment on the occurrence and intensity of highly inefficient SMBH activity remains unclear}. 

Low-level accretion-powered activity in nearby galaxies may be detected efficiently in sensitive, high spatial resolution X-ray observations, provided that one accounts properly for contamination from bright X-ray binaries to the overall nuclear emission. This issue of contamination can be substantially alleviated by restricting the search for highly inefficient nuclei to (i) early-type galaxies, as they conveniently avoid the brighter high-mass X-ray binaries produced concurrently with star formation, as well as (ii) within $\sim$30 Mpc, such that the average stellar mass enclosed within the instrument Point Spread Function (PSF) allows for a low probability of contamination from low-mass X-ray binaries\footnote{While the normalization of the X-ray luminosity function for high mass X-ray binaries scales with star formation rate, that of low-mass X-ray binaries scales with stellar mass \citep{grimm03,gilfanov04}.} \citep{gallo10,miller12b}.

\par The properties of early-type galaxies in sparsely populated regions are distinct from their counterparts in denser environments: relative to cluster/group sources, field early-type galaxies face reduced ram pressure stripping  \citep{gavazzi10,shin12} and, on average, contain more cold gas and tend to have younger stellar populations (e.g., \citealt{thomas05}). 
Proposed mechanisms for inhibiting star formation within clusters/groups include gas removal (e.g., starvation through ram pressure stripping, tidal stripping, or thermal evaporation; \citealt{treu03, moran07}) or morphological quenching (i.e., stabilization of a gas disk through the build-up of a stellar spheroid; \citealt{martig09}). Such processes would operate at low efficiency within the field.
More recent studies find that the fractional rate for AGN selected by X-ray luminosity is similar between field and cluster samples \citep{martini07,haggard10}. 
However, such investigations are concerned with {bona fide} AGN, i.e., with nuclear X-ray luminosities in excess of around $10^{43}$ \es. 
In a first attempt to bridge the gap between AGN and formally inactive galaxies, and characterize SMBH activity down to Eddington ratios as low as 10$^{-7}$, and also as a function of environment, our group carried out the AMUSE-Virgo and AMUSE-field~\cxo\ surveys\footnote{AMUSE: AGN Multi-wavelength Survey of Early-Type Galaxies}. Together, these surveys include 203 optically selected local early-type galaxies, spanning 4 decades in mass, that are {unbiased with respect to nuclear properties} \citep{gallo08, gallo10, miller12a, miller12b, leipski12, baldassare14, plotkin14, miller15}. 
These programs demonstrated a clear difference in the incidence and intensity of nuclear X-ray activity between galaxies living in isolation vs. a dense cluster such as Virgo: {after accounting for the different distributions in host galaxy stellar mass, $32\pm 6\%$ of the Virgo targets  are found to host an active nuclear SMBH \citep{gallo10}, vs. $50 \pm 7\% $ for the field \citep{miller12b}. Also, the field early-types possess greater nuclear X-ray luminosities, at a given host stellar mass, than their cluster counterparts \citep{miller12a}. 
Additionally, a Bayesian analysis, which includes luminosity upper limits, shows that, for both samples, the average nuclear X-ray luminosity scales with host stellar mass as \mstar$^{+0.8\pm 0.1}$. Consequently,  the average specific accretion rate $\langle$\lx/\mstar $\rangle$ scales with stellar mass as \mstar$^{-0.2 \pm 0.1}$, implying that black holes in lower mass galaxies tend toward higher \lx/\mstar\ ratios compared to higher mass ones \citep{miller12a}. 
This is {\it not} consistent with the {\it uniform} specific Eddington ratio distribution found for AGN \citep{aird,bongiorno12} and implies that, in the local universe, SMBH feeding is relatively more abundant/efficient at the low luminosity end of the distribution.

\par Combined, the X-ray and optical diagnostics point toward the following scenario: Small amounts of gas, possibly from residual star formation, can continue to fuel a SMBH within the nuclei of isolated field galaxies well after the assembly of the major nuclear star cluster mass, whereas gas is more likely to be depleted from dense cluster members \citep{baldassare14}. A remaining uncertainty lies in which environmental effect is primarily responsible for depleting gas: ram pressure stripping or galaxy-galaxy interactions.
At an average distance of 20 Mpc \citep{blakeslee09}, the Fornax cluster offers the only opportunity to test and {discriminate} between different environmental effects on low level SMBH accretion; at the same time, Fornax is more representative of groups and poor galaxy clusters, where the {majority} of galaxies reside. 
In fact, Virgo and Fornax show several important differences that warrant comparisons. Fornax is more regular in shape (e.g., \citealt{scharf2005}), and likely more dynamically evolved, than its northern counterpart. It is also substantially smaller and denser, with a total mass of $\simeq 7\pm 2 \cdot 10^{13}$\msun~ (vs. $4-7\cdot10^{14}$ for Virgo), a core radius of 0.4 Mpc ($\sim$ 40$\%$ that of Virgo), and a central galaxy density twice as large. The intra-cluster temperature is about twice as high in Virgo ($k$ $T\simeq 2.6$ keV) than it is in Fornax ($k$ $T\simeq 1.2$ keV). 

If ram-stripping (galaxy-gas) were primarily responsible for the lower X-ray active fraction and average nuclear X-ray luminosity in Virgo as compared to the field sample, then we would expect the active fraction and average X-ray luminosity of Fornax galaxies to be higher than in Virgo (and yet lower than the field).  \textbf{This is because ram pressure scales with the product $\rho\times \sigma^2$, where $\rho$ is the intra-cluster gas density and $\sigma$ is the cluster velocity dispersion; the central gas density and velocity dispersion of Virgo are about four and two times higher, respectively, than Fornax \citep{paolillo2002,scharf2005}, making Virgo about 15 times more effective at ram pressure stripping compared to Fornax. In contrast, galaxy-galaxy deceleration induced by dynamical friction scales as $n/\sigma^2$, where $n$ is the galaxy number density.  Since $n$ is twice as high for Fornax \citep{davies2013}, the effect of dynamical friction is expected to be a factor of about $10$ higher in Fornax than it is in Virgo. }

This paper is structured as follows: In \S~2 we describe the Fornax sample and \cxo\ data reduction procedures. Results from the X-ray data analysis are presented in \S~3, including a quantitative assessment of the low-mass X-ray binary contamination to the nuclear X-ray emission (\S~3.1), and a linear regression analysis to probe the scaling of nuclear X-ray luminosity with host stellar mass (\S~3.2). In \S~4, we carry out a mass-weighted comparison between the Fornax and Virgo/field sample, and summarize our results and conclusions in \S~5. 

\begin{table*}[t]
\begin{center}
\caption{Fornax galaxy properties \label{tab:fornax}}
\begin{tabular}{ccccccccccccccccccccc}
		\hline
		\hline  
		\multicolumn{1}{c}{ID} & \multicolumn{1}{c}{Name}  &  \multicolumn{1}{c}{Other} &\multicolumn{1}{c}{ObsID} & \multicolumn{1}{c}{D} & \multicolumn{1}{c}{Exposure}  & \multicolumn{1}{c}{log($M_{\star}$/$M_{\odot}$)} & \multicolumn{1}{c}{Count rate} & \multicolumn{1}{c}{log$L_{\rm X}$} \\
	 & & & & \multicolumn{1}{c}{(Mpc)}  & \multicolumn{1}{c}{(ks)}   & \multicolumn{1}{c}{}& \multicolumn{1}{c}{(ks$^{-1}$)} & \multicolumn{1}{c}{(erg s$^{-1}$)} \\
	\multicolumn{1}{c}{(1)} & \multicolumn{1}{c}{(2)} & \multicolumn{1}{c}{(3)} & \multicolumn{1}{c}{(4)}  & \multicolumn{1}{c}{(5)} & \multicolumn{1}{c}{(6)} & \multicolumn{1}{c}{(7)} & \multicolumn{1}{c}{(8)} & \multicolumn{1}{c}{(9)}  \\
	\hline
	1		& FCC 21*       &	NGC1316 	&	2022       	& 	21.0    	&	24.36	& 11.901	& 0.77 & 38.37 			\\
  	2   	& FCC 213*       & 	NGC1399	& 	14527  	    &	20.9		&	27.80	& 11.66 		& 0.27 & 38.15 				\\
 	3      	& FCC 219*   	& 	NGC1404 	& 	17541	 	&	20.2	 	&	24.74  	& 11.234	&	 1.57 & 38.89 				\\
	4      	& NGC 1340 	    & 	NGC1344	& 	11345	 	&	20.9		&	2.92 	& 10.956	&$<$0.33 & $<$38.84 					\\
	5   	& FCC 167*      	& 	NGC1380	& 	9526	    &	21.2		&	40.05 	& 11.071	& 0.38 & 38.19  		\\
  	6    	& FCC 276*       &	NGC1427	& 	4742   	 	&	19.6	 	&	50.05   & 10.726	& $<$0.30 & $<$37.98 				\\
  	7    	& FCC 147	    & 	NGC1374 	& 	18123	 	&	19.6		&   4.85	& 10.683	& $<$1.34 & $<$38.86 			\\
	8    	& IC 2006	    & 	IC2006 	& 	18124	 	&	20.2		& 	4.55 	& 10.505&	0.66 & 38.57 			\\
	9   	& FCC 83        &	NGC1351	&   18125		&	19.1	 	&	4.85	& 10.498	& $<$1.35 & $<$38.84 			\\
  	10   	& FCC 184*       &	NGC1387	& 	4168	 	&	19.3		& 	45.63 	& 10.922 &	0.39 & 38.58 				\\
  	11   	& FCC 63	    &	NGC1339	& 	18126	 	&	19.7		& 	4.85 & 10.355 &  2.14 & 39.06 		\\
	12  	& FCC 193*	    &  	NGC1389	&	4169		&  	21.2		& 	41.25 	& 10.482 	&$<$0.30 & $<$38.19 		\\
  	13   	& FCC 170*	    &	NGC1381	&	4170	 	& 	21.9		&	42.02	& 10.399 &	$<$1.28 & $<$38.95 				\\
  	14   	& FCC 153	    &	IC0335		&	18127	 	& 	20.8 		&	4.83 	& 9.95 	 & 0.28 & 38.89 					\\
	15  	& FCC 177	    &	NGC1380A	&	18128   	&	20.0		&	4.92	& 9.892 &	 0.62 & 38.53 				\\
  	16   	& FCC 47    	&	NGC1336	&	18129	 	& 	18.3		& 	4.85 	& 9.964 &	 $<$0.65 & $<$38.48 			\\	
  	17   	& FCC 43	    &	IC1919		&   18130	 	& 	19.8		& 	4.85    & 9.665 &	$<$0.65 & $<$38.54 				\\
	18   	& FCC 190*	    &	NGC1380B	&	4170	 	& 	20.3		&	42.02   & 9.953  &	 $<$0.27 & $<$38.15 				\\
  	19   	& FCC 310	    &	NGC1460	&	18131	 	& 	19.9		& 	4.85	& 9.924 &	 $<$1.01 & $<$38.74 			\\
	20  	& FCC 249	    &	NGC1419	&	18132		&	22.9		& 	4.84 	& 10.048 &	 0.49 &39.03 			\\
  	21   	& FCC 148	    &  	NGC1375	&	18133	 	& 	19.9 		& 	4.85	& 9.931 & 2.66	 & 39.16 			\\
	22  	& FCC 255       &	ESO358-G050	&	18134   	&	20.0		& 	4.55    & 9.644 &	 $<$0.70 & $<$38.58 				\\
  	23   	& FCC 277	    &	NGC1428 	&	18135	 	& 	20.7  	& 	4.85 	& 9.819 &	 $<$0.65 & $<$38.58 				\\
  	24   	& FCC 55	    &	ESO358-G006	   	&	18136	 	& 	20.9 		&	4.61	& 9.656  &	 $<$0.69 & $<$38.61 				\\
	25  	& FCC 152	    &	ESO358-G025		&	18137		& 	18.7 		&	4.85    & 9.444 &	 $<$0.65 & $<$38.49 				\\
	26   	& FCC 301	    &	ESO358-G059		&	18138	 	&	19.7		&	4.81 	& 9.592 &	 $<$1.02 & $<$38.73 				\\ 
	27  	& FCC 143	    &	NGC1373 	&	18139		&	19.3		&	4.85 	& 9.644 &	 $<$1.71 & $<$38.94 					\\
	28  	& FCC 182	    &	MCG-06-09-008 	&	4168		&	19.6		&	45.63	&	9.365   & 	$<$0.27 & $<$38.12 			\\
  	29  	& FCC 202	    &	NGC1396   &	4172	 	&	20.1		&	44.51	&	9.005 &	 $<$0.12 & $<$37.77 \\
	\hline
\end{tabular}
\end{center}
\medskip
Note. --  Columns: (1) Galaxy number; (2) FCC, NGC, or IC identifier. The `*' symbol indicates the use of a 2-8 keV  energy filter, as opposed to the default 0.5-8 keV used otherwise; (3) Alternative NGC, IC, ESO, or MCG identifier; (4) \cxo\ observation identity; (5) Distance \citep{blakeslee09}; (6) Sum of good time intervals; (7) Galaxy stellar mass; (8) \textsc{srcflux} generated count rate, with upper limits quoted at the 95\% confidence level; (9) 0.3-10 keV X-ray luminosity.  \\
\mbox{}
\end{table*}

\section{Sample Selection and Data Analysis} \label{sec:data}

Similarly to the Virgo and field \cxo\ surveys, a sample of Fornax early types that is unbiased with respect to nuclear properties is necessary to carry out a comparative X-ray based investigation. The Fornax Cluster Survey\footnote{\url{www.astrosci.ca/users/VCSFCS/FCS$\_$Gallery.html}} (hereafter ACS-FCS) is a \textit{Hubble Space Telescope} (HST) Advanced Camera for Survey program that targets an unbiased sample of 43 early type Fornax members spanning between $9.4 \simlt B_T \simlt 15.5$, corresponding to a factor of 275 in blue luminosity \citep{jordan2007}. 
It is constructed from the Fornax Cluster Catalog (340 members over 40 deg$^2$) by selecting morphological types E, S0, SB0, dE, deN, dS0 and dS0N down to a limiting $B_T$-magnitude brighter than 15.5. With the goal to characterize direct environmental effects (galaxy-galaxy vs. galaxy-gas) in SMBH accretion to the same depth as in Virgo, we acquired \cxo\ imaging observations of a sub-sample of the FCS targets. 

\textbf{Measurements by the ACSFCS group \citep{jordan2007} were used to estimate synthetic Vega B-band magnitudes of the host galaxies (since the synthetic Vega B-band is close in wavelength to the $g_0$ band, the
transformation introduces only a minimal uncertainty of order $0.02$ mag at most). Stellar masses are then estimated from the $g_0$ and $z_0$ band AB model magnitudes following Table 7 in \cite{bell03}.  }

\par \cxo\ ACIS imaging data are available for 29 out of the 43 early type galaxies that compose the FCS sample: 17 were acquired as part of a dedicated Cycle 17 program (17700126; PI: Gallo) where each galaxy was observed for $\sim 5$ ks, corresponding to an X-ray sensitivity threshold of $\sim 5 \times 10^{38}$ \es (we shall hereafter refer to these short, Cycle 17 pointings as ``snapshot" observations), and 12 galaxies were covered in archival \cxo\ observations, in which they are within 10 arcminutes from the aimpoint. For cases in which a galaxy was covered by multiple observations, we prioritize those observations where (i) the galaxy optical center was closest to the aimpoint, and (ii) with the lowest exposure time, so as to match the sensitivity limit of the Virgo and field surveys (which enables us to probe nuclear X-ray activity down to   \textbf{a few $10^{38}$ erg \se}, i.e. the Eddington limit for a $\sim$5 \msun~ black hole). The properties of the sample are summarized in Table \ref{tab:fornax}. 

\par Using the \cxo\ Interactive Analysis of Observation (\textsc{CIAO}) software, version 4.9, and the calibration database version 4.7.8, we analyze the observations in a systematic fashion following the general approach outlined in \cite{gallo08} and \cite{miller12b}, modulo an updated procedure for determining the optimal aperture size for measuring the flux (see below). For the 17 galaxies with snapshot observations, we first create new level 2 event and bad-pixel files, and filter the energy to within 0.5-8 keV where the ACIS is best calibrated and the background contribution is minimal.  An initial X-ray source list is created for each observation by running the Mexican-Hat wavelet source detection tool \textsc{wavdetect}, using the default significance threshold of $10^{-6}$ (corresponding to roughly one spurious detection per chip), and wavelet pixel scales of 1, 1.414, 2, 2.828, and 4. 
\par A translational astrometric correction is applied by cross-referencing the identified sources with the USNO-B1 catalog using the \textsc{CIAO~wcs\_match} script. The resulting transformation matrices are used to update the aspect solution files and the coordinate parameters in all the event files. Three observations lack sufficient matches for a shift computation and are not corrected. 

\par We search the background light curve for periods of anomalously high activity using the \textsc{CIAO~deflare} script, with 200 s binning. The script makes use of an iterative sigma-clipping algorithm that identifies background intervals $>3\sigma$ off the mean value. A ``good time'' interval is created for each observation, and visually inspected for possible artifacts, before being applied as a filter to the level 2 event file. We then generate an updated point-source list, rerunning \textsc{wavdetect} on the filtered event file, and search for an X-ray source at a position consistent with the optical center coordinates of the galaxy.  
 
\par For the 12 galaxies with archival data, the same general data reduction procedures are employed. However, for galaxies with stellar masses above $10^{10}$ \msun, we limit our analysis to the energy range 2-8 keV, 
so as to minimize diffuse free-free emission from hot gas (at the temperature and luminosity of the Fornax cluster, the contribution from thermal emission above 2 keV in the aperture sizes that we use is negligible).

\par We then use the \textsc{CIAO} \textsc{srcflux} script to compute net count rates and convert them into fluxes ($f_X$), with an automatic aperture correction. We also devise and adopt a new set of criteria to minimize contamination from an unresolved ``glow'' of low-mass X-ray binaries to the nuclear X-ray signal -- as opposed to X-ray emission from a single, centrally peaked X-ray source. The \textit{Chandra} on-axis PSF has an approximately 80\% encircled energy fraction at r$<$0.5\arcsec; a nuclear point source would therefore have the majority of its counts come from within this region. Thus, for each nuclear detection reported by \textsc{wavdetect}, we choose different apertures depending on the number of counts detected within 0.5\arcsec\ of the nominal galaxy center position, as follows: case (i) $\ge$10 counts; case (ii) $<$10 counts; case (iii) no statistically  significant detection within r=0.5 \arcsec. 
In the first case, we use a radius  r$\le$0.5\arcsec, yielding an error of $\lesssim$30\% in $f_X$ compared to the case where $f_X$ is measured from a (typical) 2\arcsec\ radius extraction region. In the second case, we use a circular region with radius r$=$2\arcsec.
In the case where the number of expected counts from low mass X-ray binaries (estimated using the luminosity function outlined in \S\ 3.1) within a 2\arcsec\ radius exceeds the measured number of counts, we accept the larger error and use again a r$=0.5$\arcsec\ radius extraction region, to minimize the contribution from the glow. In the third case, we measure $f_X$ from the smallest aperture that gives a statistically significant detection. Background counts are extracted from a source-free (i.e., with off nuclear X-ray sources, when present, masked off) annulus of inner and outer radii of 20\arcsec\ and 30\arcsec\, and centered on the X-ray source centroid position in the event of a detection, and/or the nominal galaxy optical center position in the case of a non detection. For the purpose of converting count rates into (0.3-10 keV) fluxes, we adopt an absorbed power law model with photon index $\Gamma=2$ and use \textsc{prop\_colden} to estimate the neutral hydrogen-equivalent column density for each galaxy. For the 12 archival galaxies, because many observations are off-axis, only those in which the galaxy of interest is close to the aimpoint (FCC 21, 213, 219, 167) are used in applying the aperture protocol above. For the remaining archival detection (FCC 184), a default aperture of 2\arcsec\ is used.
X-ray luminosities/limits are listed in Table \ref{tab:fornax} for the whole sample, whereas Table \ref{tab:det} gives the coordinates of the detected nuclear X-ray sources. 
\textbf{In line with previous work by our group \citep{gallo10,miller12b,miller12a}, errors are taken to be 0.1 on log\lx~and 0.1 on log\mstar/\msun, in both cases dominated by the uncertainty in the distance rather than measurement error. }

\section{Fornax Nuclear X-ray Emission Results}

\subsection{Active fraction}

The simplest estimate of the active fraction is the fraction of galaxies with nuclear X-ray sources, which corresponds to 38\%\ for the Fornax galaxies (11 out of 29). The detection rate for the snapshot survey is 6/17 (35\%), whereas among the archival data it is 5/12 (42\%). The slightly higher detection fraction for archival data could be due to the higher sensitivity (as most archival detections have luminosities below the snapshot sensitivity), but selection bias must also be important, since several of the observations are of well known AGN \textbf{(such as NGC~1316=FCC21, which is Fornax~A, and NGC~1399=FCC213, which is the central galaxy)}\footnote{Per our assessment, both are associated to low probability AGN. This is due to a combination of the (i) large enclosed nuclear masses and (ii) low X-ray luminosity values, in both galaxies. Combined, these yield a high LMXB contamination fraction probability. We stress that, for both galaxies, strong evidence for an actively accreting SMBH comes from diagnostics other than X-rays. }

\par There is a larger detection fraction for more massive galaxies, with a nuclear source detected in 8/14 galaxies with stellar mass $>10^{10}$\msun. The average nuclear \ellx=$\log$[\lx/(\es)] in the archival and snapshot samples are $38.3$ and $38.9$, respectively, and this is likely a result in the difference in sensitivity, as 4/5 of the detected archival sources have luminosities below the snapshot sensitivity. The above values, however, refer to the measured X-ray luminosities under the implicit assumption that they trace low level accretion-powered emission form a SMBH. We examine the validity of this assumption below.  

\subsection{LMXB contribution assessment}\label{sec:lmxb}

\par With regards to the origin of the nuclear X-ray signal, it is highly unlikely that any detected source is a coincident background source. The log$N$-log$S$ distribution given by \cite{morreti03} implies a negligible probability of chance coincidence, with values between $10^{-4}$ and $10^{-5}$. X-ray emission arising from stellar tidal disruption within the nuclear region is another possible origin \citep{auchettl17}, though because of its scarcity, given our sample size its influence on our analysis is negligible. Thus, we focus on quantifying the probabilistic contribution to the nuclear X-ray signal from X-ray binaries,
assessing the nature of the detected nuclear X-ray emission on a case-by-case basis (for reference, 0.5\arcsec, i.e.the \cxo\ PSF FWHM, corresponds to about 50 pc at an average distance of 20 Mpc). 

\par Both the number and cumulative luminosity of low-mass X-ray binaries (LMXBs) scale proportionally with their host galaxy's stellar mass, \mstar\ (\citealt{grimm02,gilfanov04,humphreybuote}), while, for high mass X-ray binaries, they scale with the galaxy's star formation rate (\citealt{grimm03,lehmer10, mineo12}). The star-formation rate in early-type galaxies, while poorly constrained, is undoubtedly very low, so we can safely ignore the contribution from the latter, and focus on quantifying the LMXB contribution to the nuclear signal. 

The differential X-ray luminosity function (XLF) of LMXBs in early types is well represented by a broken power law of the form $dN/dL = K_i L^{-\alpha_i}$, where $i=1,2,3$ and $\alpha_i$ has different (and progressively steeper) values for three different luminosity intervals in the range 37.3$<$\ellx$<$40.7, and $K_i$ is the normalization constant (cf. \citealt{gilfanov04}).

\begin{table}[t]
\begin{center}
\caption{Fornax nuclear X-ray sources \label{tab:det}}
\begin{tabular}{lllc}
	\hline
	\hline 
	\multicolumn{1}{c}{Galaxy name} &  \multicolumn{1}{c}{$\alpha$ (J2000)} & \multicolumn{1}{c}{$\delta$ (J2000)} & \multicolumn{1}{c}{$P_{\rm SMBH}(\%)$} \\
	\multicolumn{1}{c}{(1)}  & \multicolumn{1}{c}{(2)}  & \multicolumn{1}{c}{(3)} & \multicolumn{1}{c}{(4)} \\
	\hline
FCC 21		&	03:22:41.73	&	$-$37:12:28.55	&		0.00	\\
FCC 213		&	03:38:28.97	&	$-$35:27:02.25	&		47.1	\\
FCC 219		&	03:38:51.94	&	$-$35:35:39.40	&	92.8	\\
FCC 167		&	03:36:27.58	&	$-$34:58:34.70	&	72.2	\\
ICC 2006	&	03:54:28.41	&	$-$35:58:01.70	&		85.6	\\
FCC 184		&	03:36:57.07	&	$-$35:30:23.60	&	0.77	\\	
FCC 63	    &	03:28:06.57	&	$-$32:17:09.17	&		93.2	\\
FCC 153		&	03:35:01.06	&	$-$34:26:49.10	&	98.8	\\
FCC 177		&	03:36:47.55	&	$-$34:44:22.60	&	94.3	\\
FCC 249	    &	03:40:42.06	&	$-$37:30:39.55		&	96.4	\\
FCC 148		&	03:35:16.79	&	$-$35:15:56.40	&	99.0	\\

	\hline
\end{tabular}
\end{center}
\medskip
Columns: (1) FCC or IC Galaxy name; (2) and (3): R.A. and Dec.; (4) Probability of X-ray emission originating from SMBH-powered accretion, calculated in \S 3.2.
\end{table}

\par In line with previous works (\citealt{foord17} and references therein), for each galaxy we start by normalizing the LMXB XLF to the stellar mass enclosed within the nuclear region, $M_{\star,\rm n}$. We operate under the assumption that the stellar mass content is traced by the galaxy optical surface brightness, which we compute by performing standard aperture photometry on the publicly available FCS ACS images in the F850LP band. For the purpose of determining the radial extent of the nuclear region, we follow the same criteria as described above for the nuclear X-ray source count extraction region (i.e.,  the nuclear region here is determined by the aperture size used in \S 2; for case (ii), however, an aperture radius of 0.5\arcsec\ is used). We find nuclear fractional luminosity values ranging between $2.5\%$ (FCC 43) and $48\%$ (FCC 55). For each galaxy, the nuclear stellar mass $M_{\star,\rm n}$ is estimated by multiplying the total stellar mass by the fraction of light from the nucleus, assuming a constant mass-to-light ratio across the galaxy. New XLF normalization constants $K_i^*$ are then calculated for each energy interval based on the enclosed stellar mass. 

\par The expected average total number of LMXBs within each galaxy nucleus, and their average total luminosity, can be calculated from this re-normalized XLF as $\langle N_{\rm LMXB,tot}\rangle$ = $\int (dN/dL) dL$ and $\langle L_{\rm LMXB,tot}\rangle$=$\int L(dN/dL) dL$, respectively. These expected average values, along with the scatter in the XLF relation and the error in the measurement, can be used to determine the XRB contamination probability, $P_{\rm LMXB}$.
In general, sources brighter than $\langle L_{\rm LMXB,tot}\rangle$ are likely to be SMBHs (this is especially true above \ellx=38.67, where the XLF becomes very steep). However, because XRBs are Poisson-distributed, their luminosity may differ significantly from $\langle L_{\rm LMXB,tot}\rangle$, especially if $\langle N_{\rm LMXB,tot}\rangle$ is less than 1. More specifically, for low stellar masses, the mean and median \lx\ arising from LMXBs can differ significantly, and there is a higher chance of detecting an XRB for a given sensitivity threshold than expected from $\langle L_{\rm LMXB,tot}\rangle$. 
To account for this discrepancy, we build normalized probability distributions of the expected total luminosity in LMXBs for each galaxy, $L_{\rm LMXB,tot}$, from which the probability $P_{\rm LMXB}$ can be derived. For each galaxy containing a nuclear X-ray detection, we simulate the number of LMXBs detected above our sensitivity threshold $10^6$ times, by drawing $10^6$ Poisson deviates ($N_{1,i}$, $N_{2,i}$, and $N_{3,i}$) from the three intervals of the re-normalized XLF. We assign a luminosity to each LMXB (for a total of $N_{1,2,3,i}$ different luminosities) through inverse transform sampling the normalized XLF, then sum to obtain $L_{\rm LMXB,tot}$. The distribution of $L_{\rm LMXB,tot}$ then allows $P_{\rm LMXB}$ to be computed for each galaxy. 

\par By definition, the probability that the nuclear X-ray emission arises from a SMBH is $P_{\rm SMBH}$=(1-$P_{\rm LMXB}$), given in Table \ref{tab:det}. Roughly half of all detections (6/11) have a $P_{\rm SMBH}$ value above 90\%. Two galaxies, FCC 21 and FCC 184, have values below 1\% (0.00 and 0.77 \% respectively). Notably, FCC 21, the most massive galaxy in the sample and known AGN, has its nuclear X-ray signal ruled out from SMBH accretion entirely. 
This is likely due to the strict criteria we employ to measure the flux (\S2), and highlights the conservative nature of our approach, which allows us to provide robust lower limits on the AGN fraction. 

\begin{deluxetable}{p{68pt}llccc}
\tablecaption{Regression analysis\label{tab:fits}}
\tabletypesize{\footnotesize}
\tablewidth{8.3cm}

\tablehead{ 
 \colhead{Sample\tablenotemark{a}} & \colhead{$n$} & \colhead{$n_{\rm
 det}$} & \colhead{$\alpha$} & \colhead{$\beta$} & \colhead{${\sigma}_{\rm 0}$}}

\startdata   
Fornax ({\it meas}) 	    & 29    & 11 & $-0.67^{+0.26}_{-0.41}$ 	& $0.37^{+0.33}_{-0.25}$ 	& $0.76^{+0.33}_{-0.20}$ \\
Fornax ({\it lmxb})	    & 29   &  7.8 & $-1.18^{+0.51}_{-1.05}$ 	& $0.19^{+0.55}_{-0.50}$ 	& $1.23^{+0.93}_{-0.44}$\\[+5pt]
Virgo ({\it meas}) 	    & 100   &29	& $-0.64^{+0.12}_{-0.15}$ 	& $0.66^{+0.13}_{-0.11}$ 	& $0.62^{+0.10}_{-0.09}$ \\
Virgo ({\it lmxb}) 	    & 100   & 26.4	& $-0.72^{+0.15}_{-0.19}$ 	& $0.68^{+0.15}_{-0.13}$ 	& $0.68^{+0.13}_{-0.10}$ \\
Virgo ({\it weigh}) 	& 29    &11   .6	& $-0.53^{+0.26}_{-0.47}$ 	& $0.71^{+0.45}_{-0.35}$ 	& $0.69^{+0.37}_{-0.21}$ \\
Virgo ({\it lmxb-weigh})	& 29    & 10.6  & $-0.67^{+0.35}_{-0.89}$ 	& $0.73^{+0.59}_{-0.40}$ 	& $0.83^{+0.67}_{-0.29}$\\[+5pt]
field ({\it meas}) 	    & 103 	&52	& $-0.22^{+0.11}_{-0.13}$ 	& $0.77^{+0.11}_{-0.09}$ 	& $0.78^{+0.04}_{-0.08}$ \\
field ({\it weigh}) 	& 29    &  21.4 & $-0.07^{+0.17}_{-0.19}$ 	& $0.77^{+0.32}_{-0.29}$ 	& $0.65^{+0.19}_{-0.15}$ \\
\enddata

\tablecomments{The dependence of X-ray luminosity upon host stellar mass is parametrized as $(\text{log}L_{\rm X} -38.62) = \beta \times (\text{log} M_{\star} - 9.96 ) + \alpha$, with intrinsic random scatter ($\sigma_0$) included. The reported parameters are medians of the posterior
distributions, and errors are quoted at the 1$\sigma$ confidence level.}

\tablenotetext{a}{The various samples are defined as follows: {\it
meas\/}: measured \lx\ values are employed in the regression analysis, regardless of the estimated LMXB contamination; {\it lmxb\/}: detections probabilistically treated as an upper limit depending on the estimated $P_{\rm SMBH}$; {\it weigh\/}: Mass distribution-weighted comparison, where a sub-sample of 29 objects (i.e., same size as the Fornax sample) has been drawn from the larger sample according to the weighing functions shown in Figure \ref{fig:fits}; {\it lmxb-weigh\/}: Mass distribution-weighted comparison, with detections being probabilistically treated as an upper limit within each sub-sample. Note that the {\it lmxb\/} and {\it lmxb-weigh\/} sub-samples are omitted from the field case because the lack of uniform HST coverage prevented us from re-assessing the LMXB contamination probabilities following the procedure outlined in \S 3.2. \textbf{For each sample (i.e., row) $n$ indicates the number of targets under consideration, i.e.: 29 for Fornax; 100 for Virgo; 103 for the field; and 29 again for either Virgo or the field once a mass-corrected comparison to Fornax is performed; $n_{\rm det}$ refers to the mean number of X-ray detections after accounting for the different mass distribution and/or LMXB contamination; the value of $n_{\rm det}$ is computed as the average over all iterations/sub-samples. }
}
\end{deluxetable}

\subsection{Dependence on host stellar mass}

\par The active fraction (for a given sensitivity) is the simplest metric of black hole activity. Motivated by prior work which established a functional relationship between the nuclear X-ray luminosity and host galaxy stellar mass, we also search for such a relation within our data; specifically, the presence of a linear relationship in logarithmic space, of the form:
\begin{equation} 
\label{eq:1}
(\text{log}L_X - 38.62) = \beta \times (\text{log} M_{\star} - 9.96) + \alpha
\end{equation} 
where variables are centered on their respective median values, and 0.1~dex uncertainties are assumed in both variables.
The regression analysis is implemented in \textsc{IDL} with the Bayesian code \textsc{linmix\_err.pro} \citep{kelly07}, which accounts for censored data points and yields reliable results with as many as 90\% of the data points having upper limits \citep{foord17}. 
We adopt the default independent variable mixture modeling, which makes use of three Gaussians, and a minimum of 10,000 iterations with Gibbs sampling. The best-fit parameters are estimated as the median values from 10,000 draws from the posterior distributions. 1$\sigma$ confidence intervals are estimated as the 16th and 84th percentiles of the normalized posterior distribution, after verifying that the posterior distributions appear consistent with a normal distribution. Results from the regression analysis are listed in Table \ref{tab:fits}. 

\par The regression analysis is first run assuming that all nuclear X-ray sources are SMBHs, adopting the \lx\ values listed in Table \ref{tab:fornax} for the FCS sample as dependent variables.
The best-fit linear relation is found to be  $({\rm log}L_X - 38.62) = -0.67^{+0.26}_{-0.41} + 0.37^{+0.33}_{-0.25} \cdot (\text{log} M_{\star} - 9.96)$. The slope $\beta = 0.37$ is inconsistent with zero at 1.5$\sigma$ (first line of Table \ref{tab:fits}). 
The Pearson correlation coefficient of 0.37 for our sample also indicates a significant correlation. However, the existence of a correlation cannot be ruled out at $>$3 sigma confidence.

\par In order to account for the LMXB contamination, we re-run \textsc{linmix\_err} 1,000 times. Within each iteration, and for each detected source, we draw a uniform deviate $U\in(0,1)$. If $U$ exceeds $P_{\rm SMBH}$, the source is then treated as an upper limit. The posterior distributions of the 1,000 runs are then combined, from which the fit parameters and errors are calculated in an equivalent manner as above (Table \ref{tab:fits}). The resulting slope, $\beta = 0.19$ (labeled as \textit{lmxb}), is consistent with a zero value within $1 \sigma$, so we cannot conclude that the X-ray luminosity is correlated with host stellar mass. The mild discrepancy between the two treatments (i.e., without and with LMXB contamination included in the fits) likely results from the linear correlation of between LMXB luminosity and stellar mass. Nonetheless, the best-fit slopes with and without correction for LMXB contamination are consistent with each other because of the small sample size.

\begin{figure}[t]
\begin{center}
\includegraphics[width=0.4\textwidth]{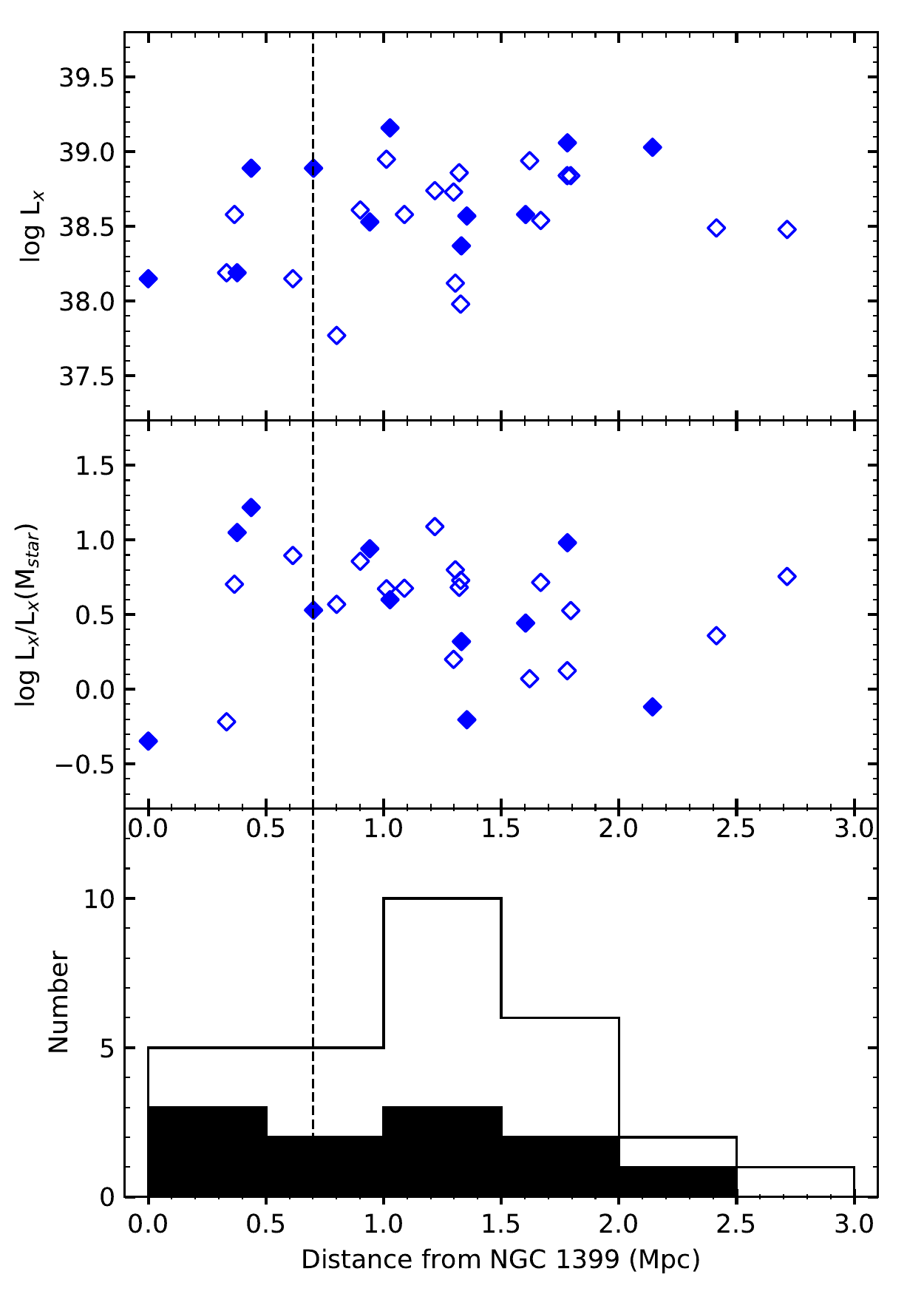}
\end{center}
\caption{Properties of the FCS galaxies in our \cxo\ sample as a function of their distance from the central cluster galaxy, NGC 1399. The vertical dotted line marks 50\% of the cluster virial radius. Filled/open symbols represent nuclear X-ray detections/upper limits.
{\it Top}: measured nuclear {X-ray}
luminosity. {\it Middle}: residual X-ray luminosity, calculated as the ratio of the measured value to the expected value from the best-fitting relation with host stellar mass. {\it Bottom}: number of targeted galaxies per (radial) distance bin; the filled histogram represents the galaxies with nuclear X-ray detections.  }
\label{fig:dist}
\end{figure}

\section{Environmental dependence of SMBH activity} 

\par Here we compare the active fraction, average X-ray luminosity, and X-ray luminosity:host stellar mass correlation between the Fornax sample and two other studies addressing sub-Eddington X-ray emission from galactic nuclei down to comparably low sensitivities, i.e. the AMUSE Virgo \citep{gallo10} and AMUSE-field \citep{miller12b, miller12a} \cxo\ surveys, including 100 early types within the Virgo cluster, and 103 early types in isolation.

\par Following \cite{miller12a}, we first consider here any possible dependence of the Fornax nuclei X-ray luminosities on radial distance, calculated from the central cluster galaxy NGC 1399.  Figure \ref{fig:dist} shows \lx, residual X-ray luminosity (i.e, with respect to the best-fitting relation with \mstar), and X-ray detection fraction as a function of the distance from NGC 1399. No statistically significant trends are observed in any three of these criteria, and thus we shall now proceed in analysis of the Fornax sample in its entirety. 
\subsection{Nuclear Activity in the Fornax, Virgo, and field samples}

For consistency, $L_{\rm X}$ values for Virgo galaxies are recalculated according to the updated photometry protocol outlined in \S~2. Revised $P_{\rm SMBH}$ values also are updated according to the LMXB contamination assessment technique described in \S~\ref{sec:lmxb}\footnote{The revised photometry and LMXB assessment technique presented in this work yield a slightly lower rate of nuclear X-ray detections (29 out of 100, as opposed to 32) and more conservative contamination probabilities compared to those quoted in the original Virgo analysis \citep{gallo10}.  Albeit superior, owing to the small number statistics, the new method presented here does not affect our conclusions from a qualitative perspective, in the sense that the revised, LMXB contamination-corrected active fraction for the Virgo sample is consistent, within the errors, with the quoted value from the original paper. }. 

\begin{figure}[t]
\begin{center}
\includegraphics[width=0.45\textwidth]{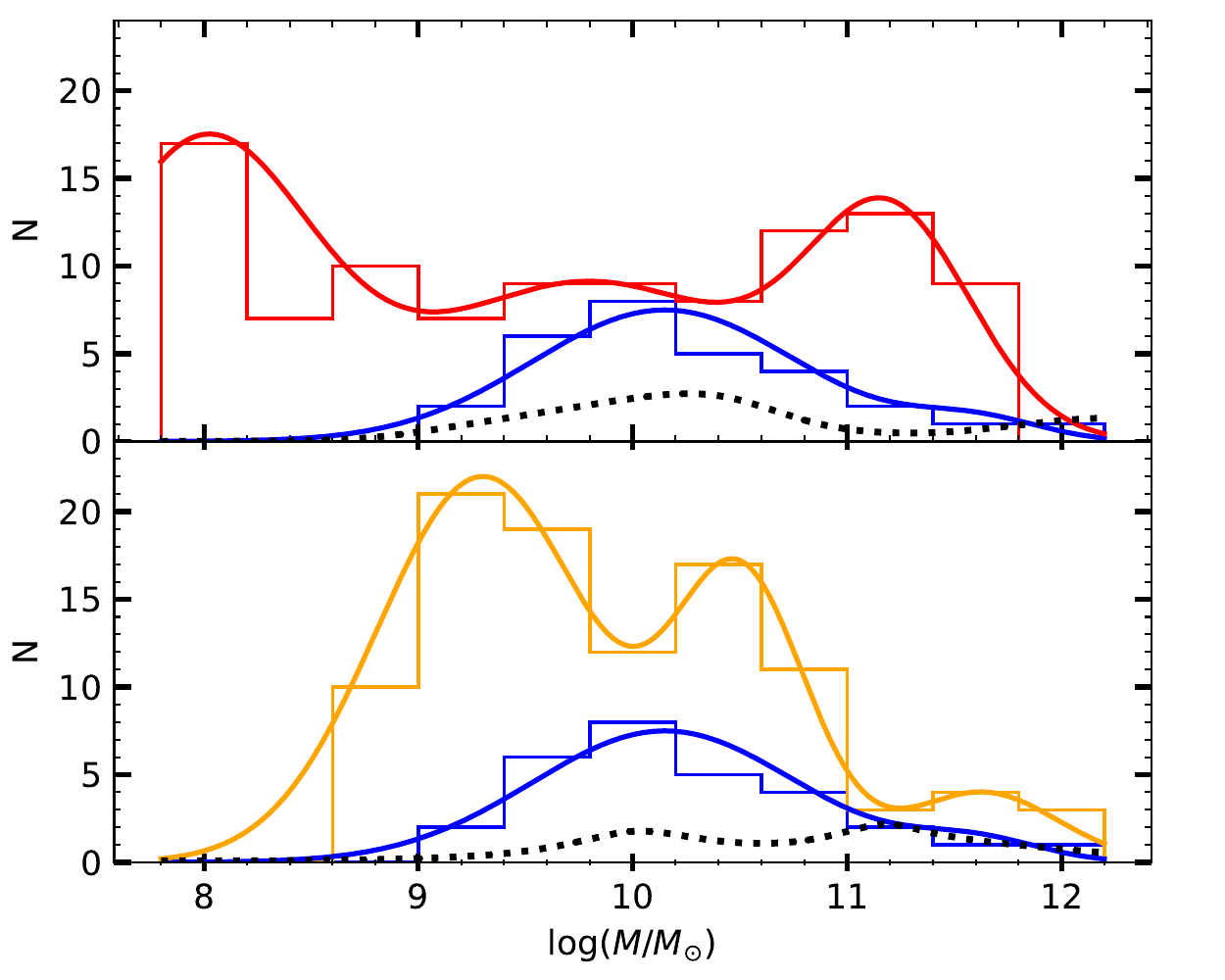}
\end{center}
\caption{Stellar mass distribution of Fornax (blue), field (red), and the corresponding weighting function (black), illustrated on the top panel. The bottom panel displays the Fornax distribution compared to Virgo (orange), similarly with the corresponding weighting function.}
\label{fig:weigh}
\end{figure}

\par Looking at the measurements at face value (labeled as \textit{meas} values in Table \ref{tab:comp}), we derive an active fraction of 29/100 = 29.0\%, and 52/103 = 50.5\% for Virgo and field respectively. The face-value Fornax active fraction of 11/29 = 37.9\% is intermediate between those of the full Virgo and field samples. 

If the active fraction in Fornax were the same as in Virgo, there would be a 2.5\% chance of detecting $\ge$11/29 galaxies, whereas if it were the same as the field, the chance for detecting $\le$11/29 galaxies would be only 0.5\%. The average detected \ellx\ are 38.79, 38.72, and 39.13 for the Fornax, Virgo, and field sample, respectively; the average error in the measured \ellx, computed by accounting for the errors on distance and detected X-ray counts, is $\pm 0.45$ dex. So whereas the average \ellx\ of Fornax and Virgo are in agreement, the field \ellx\ is marginally greater than Fornax. Considering, however, the strong dependence of nuclear activity and star formation on $M_*$ \citep{treu2005a}, we expect any comparison of active fraction and \lx\ between the samples of Fornax, Virgo, and the field to be strongly influenced by the differing $M_*$ distributions. We address this issue below.

\subsubsection{Mass and LMXB-Corrected Comparison}

\par  Here we carry out a comparison of the Virgo and field samples to Fornax while accounting for the differing mass distributions. We additionally account for the LMXB contamination on the mass-corrected Virgo sample. For the field sample, owing to the lack of uniform \hst\ coverage, we are unable to revise the LMXB contamination probabilities according to the procedure outlined above. This does not significantly affect the analysis, however, as it is the comparison between Fornax and Virgo galaxies that is of primary interest. We further emphasize that the three samples are unbiased with respect to nuclear X-ray properties and capture all SMBH activity at \ellx$>38.4$.

\par To control for the different host stellar mass distributions, we construct mass weighted sub-samples drawn from Virgo and field (100 and 103 galaxies respectively), which have a mass distribution consistent with that of our smaller Fornax sample (29 galaxies). This is accomplished through the use of two weighting functions, defined as the ratio of the Virgo/field mass distribution to that of Fornax, with the distributions estimated as the sum of multiple Gaussians (see Fig. \ref{fig:weigh}). Converted to probability density functions, the weighting functions are then used to draw 29 galaxies without replacement. A total of $10^6$ such sub-samples are drawn from the parent Virgo and field samples. The average mass, detection luminosity, and number of detections are computed from the drawn sub-samples (Fornax values taken as is), along with 25th, 50th, and 75th percentile values, taken as the closest discrete value for all three classifications. Results from this comparison are labeled as \textit{weigh} values in Table \ref{tab:comp}.

\par As in \S 3.2, we then sampled detections according to their $P_{\rm SMBH}$ value. We note that the LMXB-corrected parameters for the Fornax sample are compared heuristically by sampling the same sample multiple times, whereas for the Virgo sample $10^6$ distinctly drawn sub-samples are each individually sampled once. Results from this comparison are labeled as \textit{lmxb-weigh} values in Table \ref{tab:comp}.

\par For the Virgo and field mass-weighted samples, we approximate the probability distribution of the number of detections by centering a Poisson distribution on the mean detection value found. We measure an averaged active fraction of $40\% \pm 11.7\%$ and $74\% \pm 15.9\%$ for Virgo and field, respectively (Fornax is $38\% \pm 11.4\%$). The mass-weighted Virgo active fraction is thus consistent with that of Fornax, while the mass-weighted field sample shows an increased disagreement with Fornax in active fraction, compared to the non-weighted full field sample. When further correcting for LMXB contamination, the Virgo active fraction is $36.6\% \pm 11.2\%$, whereas Fornax's active fraction is $26.6\% \pm 9.6\%$. These values remain in tentative agreement, although the data suggest that the Fornax LMXB contamination-corrected active fraction is smaller: given an average of 7.8 detections in 29 observations, the probability of detecting $>10$ is 9.6\%, and of detecting $>11$ is 6.7\%.  

\par The average detection luminosity, $\ell_{{X,n_{\rm det}}}$, for the weighted Virgo samples is 38.73, less than that of Fornax at 38.79, as well as that of the weighted field samples at 38.98. Considering that the average error in \ellx\ of Fornax is $\pm 0.45$, however, the three samples are in agreement. These average luminosities may be misleading if they tend to come from galaxies with very different masses; however, in a randomly pulled Virgo mass-weighted sub-sample, 9 detections are recorded between $10^{10} - 10^{11} M_*$, all with luminosities between 38.4 and 38.8. In the Fornax sample, four detections are found in the same mass bin, two of which have luminosities above 39, so the apparently larger luminosities of the Fornax detections is not due to a large difference in the masses of galaxies containing detections. For the LMXB contamination-corrected, mass-weighted Virgo sub-samples, the mean $\ell_{X,n_{\rm det}}$ is 38.77, equivalent to that of the LMXB-corrected Fornax sample at 38.77. There is no evidence for any statistically significant difference between the detection luminosities between Fornax and Virgo. 

\par When constructing mass-weighted Virgo samples to match the Fornax mass distribution, we conclude that no statistically significant difference in either the active fraction or average nuclear X-ray luminosity between the two clusters is found. This is true regardless of whether or not we account for LMXB contamination, although there is a suggestion that the active fraction in Fornax may be slightly lower than in Virgo galaxies with the same mass. Further, as expected, the active fraction and average luminosity of Fornax are systematically lower than those of the field sample for both face-value and mass-weighted comparisons, though the difference is more pronounced in active fraction, whereas the difference in luminosity remains within the error. The pronounced change of the Fornax active fraction after accounting for LMXB contamination, relative to Virgo, suggests a greater amount of LMXB contamination within the Fornax sample (likely due to the larger fraction of massive galaxies with low nuclear X-ray luminosities in this sample). Assuming the LMXB-corrected and LMXB-corrected plus mass-weighted comparisons between Fornax/Virgo reflect more accurately the true active fraction within the samples, the data suggest the active fraction of Fornax to be lower.

\begin{deluxetable}{p{65pt}rrrcrr}
\tablecaption{Comparison of Sample properties\label{tab:comp}}
\tabletypesize{\footnotesize}
\tablewidth{8.8cm}

\tablehead{\colhead{Sample} & \colhead{$n$} & \colhead{Mean} &
\colhead{25th} & \colhead{50th} & \colhead{75th} }

\startdata   
\multicolumn{6}{c}{$\log{(M_{\star}/M_{\odot})}$}  \\[+1pt]
Fornax ({\it meas})  & 29      &   10.24  &  9.67  &  9.96  & 10.68  \\ [+3pt]
Virgo ({\it meas})  & 100    &    9.9  &  9.2  &  9.75  &  10.4 \\   
Virgo ({\it weigh}) & 29     &    10.24  &  9.7  &  10.2  & 10.6  \\   [+3pt]
field ({\it meas})   & 103      &    9.66  & 8.5  & 9.7  & 10.8  \\ 
field ({\it weigh})  & 29        &    10.24  & 9.7  & 10.3  & 10.7  \\ [+6pt]

\multicolumn{6}{c}{$n_{\rm det}$} \\  [+1pt]
Fornax ({\it meas})   & 29      &   11  &  -  &  -  & -  \\
Fornax ({\it lmxb})   & 29       &   7.8 &  7 &  8  & 8  \\ [+3pt]
Virgo ({\it meas})    & 100  &    29  &  -  &  -  & -  \\   
Virgo ({\it weigh})   & 29       &    11.6  &  10  &  12  & 13   \\
Virgo ({\it lmxb-weigh}) & 29    &    10.6  &  9  &  11  & 12  \\ [+3pt] 
field ({\it meas})   & 103       &    52  & -  & -  & -  \\ 
field ({\it weigh})  & 29        &   21.4  & 20  & 21  & 23  \\ [+6pt]

\multicolumn{6}{c}{$\log{(L_{\rm X, n_{\rm det}})}$} \\[+2pt]
Fornax ({\it meas})  & 29      &    38.79  &  38.57  &  38.84  & 39.03  \\	
Fornax ({\it lmxb})  & 29       &   38.77  &  38.53  &  38.89 & 39.03  \\ [+3pt]
Virgo ({\it meas})  & 100    &    38.72  &  38.5  &  38.6  & 38.8  \\   
Virgo ({\it weigh}) & 29  &  38.73  &  38.5  &  38.6  & 38.8 \\    
Virgo ({\it lmxb-weigh}) & 29   &    38.77  &  38.5  &  38.6  & 38.8  \\[+3pt]
field ({\it meas})     & 103     &    39.13  & 38.7  & 39.0  & 39.4  \\ 
field ({\it weigh})    & 29      &   38.98  & 38.6  & 38.9  & 39.3  \\ [-2pt]

\enddata
\tablecomments{The mean and 25th, 50th, and 75th percentile properties of $10^6$ mass-distribution weighted sub-samples drawn from the parent Virgo and field samples, alongside the full three samples and LMXB-corrected and weighted samples, are listed above. The sample classifications are analogous to the definitions outlined in Table 3, including the definition of $n$ and $n_{\rm det}$. field ({\it lmxb-weigh}) properties are not listed, due to the lack of reliable HST imaging from which to compute $P_{\rm SMBH}$ probabilities.}
\end{deluxetable}

\subsection{\lx:\mstar\ correlation in Fornax, Virgo, field}\label{sec:enviro}

\par Next, we compare the best-fit \lx:\mstar\ relationship for Fornax to that in the Virgo and field samples. Similarly to what we did for the active fraction and mean luminosity assessment, we first run the regression analysis on the measured (i.e., non LMXB-contamination-corrected) \lx\ values for the Virgo and the field galaxies. 
This is then compared with the results from LMXB-corrected analysis (labeled as \textit{lmxb} in Table \ref{tab:fits}) , where, for the Virgo sample, \textsc{linmix\_err} is run 5,000 times, probabilistically varying each detection as a detection/limit based on their respective $P_{\rm SMBH}$ probabilities. Best-fitting values of the intercept, slope and intrinsic scatter are listed in Table \ref{tab:fits} and labeled as \textit{meas} and \textit{lmxb} sample, respectively, for the face-value and contamination-corrected samples (Table \ref{tab:fits}). 
\par For the Fornax face-value sample, we previously established the best-fit slope as $\beta = 0.37^{+0.37}_{-0.25}$, with an intrinsic scatter $\sigma = 0.76 ^{+0.11}_{-0.09}$. The median slopes for the Virgo and field samples are $\beta = 0.66^{+0.13}_{-0.11}$ and $\beta = 0.77^{+0.11}_{-0.09}$ respectively. 

The slope of the LMXB-contamination-corrected Virgo sample remains nearly unchanged, from $\beta = 0.66^{+0.13}_{-0.11}$ to $0.68^{+0.15}_{-0.13}$, with the intrinsic scatter also increasing slightly. The Fornax slope, $\beta = 0.37^{+0.33}_{-0.25}$, is consistent with the Virgo slope within 1 $\sigma$, and differs from the field slope at 1.2 $\sigma$. The LMXB-corrected Fornax slope, $\beta = 0.19^{+0.55}_{-0.50}$, is similarly consistent with Virgo's within 1$\sigma$. 

\par To control for the different mass distributions, we re-run \textsc{linmix\_err} on $500$ (reduced from the $10^6$ samples used previously for feasibility) of the Fornax-matched Virgo and field sub-samples, and combine the posterior distributions of each sub-sample, from which we pull the median values for the intercept, slope and intrinsic scatter. Lastly, we account for LMXB contamination to the mass-weighted samples by probabilistically varying, within each drawn sub-sample, each \lx\ 10 times according to their $P_{\rm SMBH}$ probabilities, for a total of 5,000 runs. The results from this final comparison are listed in Table \ref{tab:fits} and labeled as \textit{weigh} and \textit{lmxb-weigh}, respectively, and are shown in Figure \ref{fig:fits}. 

The mass-weighted Virgo best-fit slope, $\beta = 0.71^{+0.45}_{-0.35}$, is consistent, within the errors, with the face-value slope. Similarly, the slope for mass-weighted plus LMXB-corrected Virgo sample increases slightly to  $\beta = 0.73^{+0.59}_{-0.40}$, though again, it is consistent, within the errors, with the face-value slope. The face-value Fornax slope differs from the mass-weighted Virgo slope at 1.0 $\sigma$, and from the mass-weighted, contamination-corrected slope at the 1.1 $\sigma$ level. The best-fit relations for Fornax and Virgo are shown in Figure 3 (the field sample is omitted from this comparison as the lack of uniform HST coverage prevented us from re-assessing the LMXB contamination following our new methodology, outlined in \S 3.2). 

\begin{figure}[t]
\includegraphics[width=0.5\textwidth]{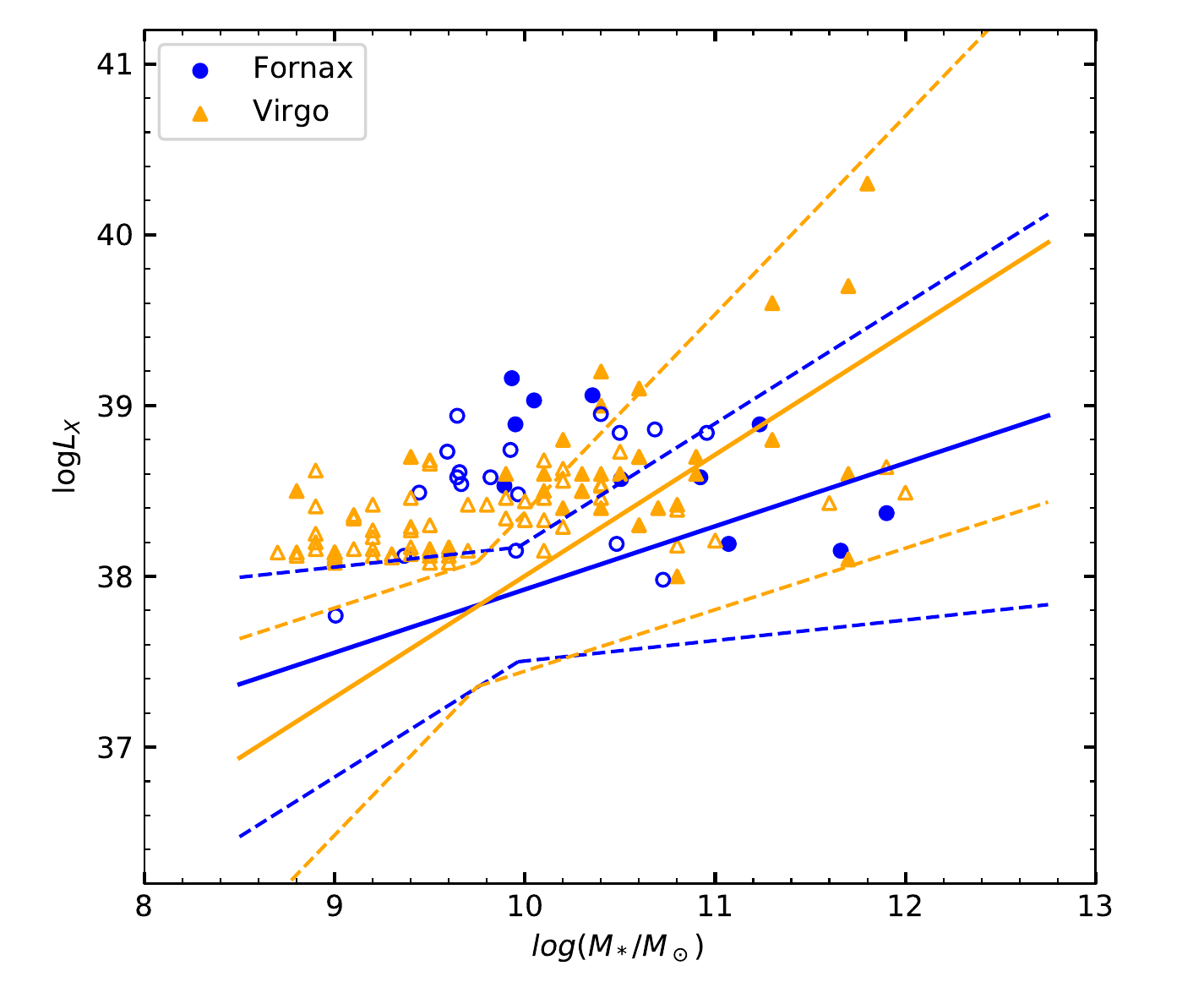}
\caption{The X-ray luminosities and upper limits (filled and open symbols, respectively) for both the Virgo (orange) and Fornax (blue) samples are shown as a function of host stellar mass. Solid lines represent the median of the slopes' posterior distributions, with 3$\sigma$ errors enclosed within the dashed lines.}
\label{fig:fits}
\end{figure}

\section{Summary and Conclusions}

\par 
Previous studies of local low-level SMBH activity, through the AMUSE-Virgo and AMUSE field surveys, demonstrate a difference in the incidence and intensity of such activities, which can be interpreted as evidence of environmental effects on SMBH fueling and accretion.  
The Fornax Cluster, being significantly denser and possessing an intra-cluster temperature roughly twice that of Virgo, offers the opportunity to further examine local low-level SMBH activity and the impact of different environmental effects. From the ACS FCS, we target 17 Fornax galaxies with the \textit{Chandra X-ray Telescope}, which we then combine with 12 existing archival observations, all of which are unbiased with respect to nuclear properties and capture all SMBH activity at \ellx$>38.4$.
The results from this paper can be summarized as follows.

\begin{enumerate}
  \item A nuclear X-ray point source consistent with the optical center of the galaxy is found in 11 out of 29 of the Fornax galaxies. Taking into consideration the contained stellar mass within the nuclear aperture of each nuclear X-ray detection as well as their respective measured luminosity, we quantify the estimated likelihood of each detection originating from SMBH powered accretion. We then compute a weighted active fraction of $26.6\% \pm 9.6\%$. 
  
  \item We employ a Bayesian regression analysis to search for a linear relationship between the nuclear X-ray luminosity and host galaxy stellar mass, and find a best-fit relation (\ellx$- 38.62) = -0.67^{+0.26}_{-0.41} + 0.37^{+0.33}_{-0.25} \cdot (\text{log} M_{\star} - 9.96)$. The large uncertainties prevent us from placing strong constraints on the \lx:$M_{\star}$ relation. When accounting for LMXB contamination for each detection in the analysis, a lower intercept and much flatter slope are returned ($\beta = 0.19^{+0.55}_{-0.50}$), due to the high chance of contamination for the higher mass galaxies in the sample. 
  
  \item We compare the incidence and intensity of nuclear activity, measured through their average \ellx\ and active fraction, between Fornax, Virgo and field galaxies. 
 After correcting for the different mass distributions, the measured Fornax active fraction is found to be less than the field fraction, at $>3\sigma$. These findings are consistent with the conclusions of \cite{miller12b}, in that lower activity is found in cluster galaxies. This is consistent with a scenario where the funneling of gas to the nuclear regions has been inhibited more effectively in a cluster environment, arguably via ram pressure stripping or galaxy-galaxy deceleration induced by dynamical friction (see also \citealt{baldassare14}).
 
 \item After accounting for the different mass distributions, the average \ellx\ and active fraction in Fornax and Virgo are consistent within the errors, although the data suggest that, after LMXB contamination corrections, the Fornax active fraction may be smaller than Virgo's. If significant, this would tentatively suggest that dynamical friction (expected to be a factor $\sim$10 higher in Fornax than in Virgo), may play a more important role in gas removal compared to ram-pressure stripping.

\end{enumerate}

\bibliographystyle{aasjournal}
\bibliography{NL}
\end{document}